\documentclass[10pt]{article}

\usepackage{amsmath}
\usepackage{amssymb}

\usepackage{graphicx}

\usepackage{cite}

\usepackage{color}

\usepackage[labelfont=bf,labelsep=period,justification=raggedright]{caption}

\makeatletter
\renewcommand{\@biblabel}[1]{\quad#1.}
\makeatother

\date{}

\begin{document}

\begin{flushleft}
{\Large \textbf{Zipf's Law Leads to Heaps' Law: Analyzing Their
Relation in Finite-Size Systems} }
\\
Linyuan L\"u$^{1,2}$, Zi-Ke Zhang$^{2}$, Tao Zhou$^{1,2,3,\ast}$
\\
\bf{1} Web Sciences Center, University of Electronic Science and
Technology of China, Chengdu 610054, PR China
\\
\bf{2} Department of Physics, University of Fribourg, Chemin du
Mus\'ee 3, Fribourg 1700, Switzerland
\\
\bf{3} Department of Modern Physics, University of Science and
Technology of China, Hefei 230026, PR China
\\
$\ast$ E-mail: zhutou@ustc.edu
\end{flushleft}

\section*{Abstract}

\textbf{\emph{Background}}: Zipf's law and Heaps' law are observed
in disparate complex systems. Of particular interests, these two
laws often appear together. Many theoretical models and analyses are
performed to understand their co-occurrence in real systems, but it
still lacks a clear picture about their relation.\\
\textbf{\emph{Methodology/Principal Findings}}: We show that the
Heaps' law can be considered as a derivative phenomenon if the
system obeys the Zipf's law. Furthermore, we refine the known
approximate solution of the Heaps' exponent provided the Zipf's
exponent. We show that the approximate solution is indeed an
asymptotic solution for infinite systems, while in the finite-size
system the Heaps' exponent is sensitive to the system size.
Extensive empirical analysis on tens of disparate systems
demonstrates that our refined results can better capture the
relation between the Zipf's and Heaps' exponents.\\
\textbf{\emph{Conclusions/Significance}}: The present analysis
provides a clear picture about the relation between the Zipf's law
and Heaps' law without the help of any specific stochastic model,
namely the Heaps' law is indeed a derivative phenomenon from Zipf's
law. The presented numerical method gives considerably better
estimation of the Heaps' exponent given the Zipf's exponent and the
system size. Our analysis provides some insights and implications of
real complex systems, for example, one can naturally obtained a
better explanation of the accelerated growth of scale-free networks.

\section*{Introduction}

Giant strides in Complexity Sciences have been the direct outcome of
efforts to uncover the universal laws that govern disparate systems.
Zipf's law \cite{Zipf1949} and Heaps' law \cite{Heaps1978} are two
representative examples. In 1940s, Zipf found a certain scaling law
in the distribution of the word frequencies. Ranking all the words
in descending order of occurrence frequency and denoting by $z(r)$
the frequency of the word with rank $r$, the Zipf's law reads
$z(r)=z_\texttt{max}\cdot r^{-\alpha}$, where $z_\texttt{max}$ is
the maximal frequency and $\alpha$ is the so-called Zipf's exponent.
This power-law frequency-rank relation indicates a power-law
probability distribution of the frequency itself, say $p(z)\sim
z^{-\beta}$ with $\beta$ equal to $1+1/\alpha$ (see
\textbf{Materials and Methods}). As a signature of complex systems,
the Zipf's law is observed everywhere \cite{Clauset2009}: these
include the distributions of firm sizes \cite{Axtell2001}, wealths
and incomes \cite{Dragulescu2001}, paper citations
\cite{Redner1998}, gene expressions \cite{Furusawa2003}, sizes of
blackouts \cite{Bai2006}, family names \cite{Baek2007}, city sizes
\cite{Cordoba2008}, personal donations \cite{Chen2009}, chess
openings \cite{Blasius2009}, traffic loads caused by YouTube videos
\cite{Abhari2010}, and so on. Accordingly, many mechanisms are put
forward to explain the emergence of the Zipf's law
\cite{Mitzenmacher2004,Newman2005}, such as the \emph{rich gets
richer} \cite{Simon1955,Barabasi1999}, the \emph{self-organized
criticality} \cite{Bak1996}, \emph{Markov Processes}
\cite{Kanter1995}, \emph{aggregation of interacting individuals}
\cite{Marsili1996}, \emph{optimization designs} \cite{Carlson1999}
and the \emph{least effort principle} \cite{Cancho2003}. To name
just a few.

Heaps' law \cite{Heaps1978} can also be applied in characterizing
natural language processing, according to which the vocabulary size
grows in a sublinear function with document size, say $N(t)\sim
t^\lambda$ with $\lambda<1$, where $t$ denotes the total number of
words and $N(t)$ is the number of distinct words. One ingredient
causing such a sublinear growth may be the memory and bursty nature
of human language \cite{Ebeling1994,Kleinberg2003,Altmann2009}. A
particular interesting phenomenon is the coexistence of the Zipf's
law and Heaps' law. Gelbukh and Sidorov \cite{Gelbukh2001} observed
these two laws in English, Russian and Spanish texts, with different
exponents depending on languages. Similar results were recently
reported for the corpus of web texts \cite{Serrano2009}, including
the \emph{Industry Sector database}, the \emph{Open Directory} and
the \emph{English Wikipedia}. Besides the statistical regularities
of text, the occurrences of tags for online resources
\cite{Cattuto2007,Cattuto2009}, keywords for scientific publications
\cite{Zhang2008}, words contained by web pages resulted from web
searching \cite{Lansey2009}, and identifiers in modern Java, C++ and
C programs \cite{Zhang2009} also simultaneously display the Zipf's
law and Heaps' law. Benz \emph{et al.} \cite{Benz2008} reported the
Zipf's law of the distribution of the features of small organic
molecules, together with the Heaps' law about the number of unique
features. In particular, the Zipf's law and Heaps' law are closely
related to the evolving networks. It is well-known that some
networks grow in an accelerating manner
\cite{Dorogovtsev2001,Smith2007} and have scale-free structures (see
for example the WWW \cite{Broder2000} and Internet
\cite{Zhang2008b}), in fact, the former property corresponds to the
Heaps' law that the number of nodes grows in a sublinear form with
the total degree of nodes, while the latter is equivalent to the
Zipf's law for degree distribution.

Baeza-Yates and Navarro \cite{Baeza-Yates2000} showed that the two
laws are related: when $\alpha>1$, it can be derived that if both
the Zipf's law and Heaps' law hold, $\lambda=\frac{1}{\alpha}$. By
using a more polished approach, Leijenhorst and Weide
\cite{Leijenhorst2005} generalized this result from the Zipf's law
to the Mandelbrot's law \cite{Mandelbrot1960} where $z(r)\sim
(r_c+r)^{-\alpha}$ and $r_c$ is a constant. Based on a variant of
the Simon model \cite{Simon1955}, Montemurro and Zanette
\cite{Montemurro2002,Zanette2005} showed that the Zipf's law is a
result from the Heaps' law with $\alpha$ depending on $\lambda$ and
the modeling parameter. Also based on a stochastic model, Serrano
\emph{et al.} \cite{Serrano2009} claimed that the Zipf's law can
result in the Heaps' law when $\alpha>1$, and the Heaps' exponent is
$\lambda=\frac{1}{\alpha}$. In this paper, we prove that for an
evolving system with stable Zipf's exponent, the Heaps' law can be
directly derived from the Zipf's law without the help of any
specific stochastic model. The relation $\lambda=\frac{1}{\alpha}$
is only an asymptotic solution hold for very-large-size systems with
$\alpha>1$. We will refine this result for finite-size systems with
$\alpha\gtrsim 1$ and complement it with $\alpha<1$. In particular,
we analyze the effects of system size on the Heaps' exponent, which
are completely ignored in the literature. Extensive empirical
analysis on tens of disparate systems ranging from keyword
occurrences in scientific journals to spreading patterns of the
novel virus influenza A (H1N1) has demonstrated that the refined
results presented here can better capture the relation between
Zipf's and Heaps' exponents. In particular, our results agree well
with the evolving regularities of the accelerating networks and
suggest that the accelerating growth is necessary to keep a stable
power-law degree distribution. Whereas the majority of studies on
the Heaps' law are limited in linguistics, our work opens up the
door to a much wider horizon that includes many complex systems.

\section*{Results}

\subsection*{Analytical Results}
For simplicity of depiction, we use the language of word statistics
in text, where $z(r)$ denotes the frequency of the word with rank
$r$. However, the results are not limited to language systems. Note
that $(r-1)$ is the very number of distinct words with frequency
larger than $z(r)$. Denoting by $t$ the total number of word
occurrences (i.e., size of the text) and $N(t)$ the corresponding
number of distinct words, then
\begin{equation}
r-1=\int^{z_\texttt{max}}_{z(r)}N(t)p(z')\texttt{d}z'. \label{eq1}
\end{equation}
Note that $p(z)=Az^{-\beta}$ with $A$ a constant. According to the
normalization condition $\int^{z_\texttt{max}}_1 p(z)\texttt{d}z=1$,
when $\beta>1$ and $z_\texttt{max}\gg 1$ (these two conditions are
hold for most real systems),
$A=\frac{\beta-1}{1-z_\texttt{max}^{1-\beta}}\approx\beta-1$.
Substituting $p(z')$ in Eq.~\ref{eq1} by $(\beta-1)z'^{-\beta}$, we
have
\begin{equation}
r-1=N(t)\left[z(r)^{1-\beta}-z_\texttt{max}^{1-\beta}\right].
\label{eq2}
\end{equation}
According to the Zipf's law $z(r)=z_\texttt{max}\cdot r^{-\alpha}$
and the relation between the Zipf's and power-law exponents
$\beta=1+\frac{1}{\alpha}$, the right part of Eq. \ref{eq2} can be
expressed in term of $z_\texttt{max}$ and $\alpha$, as
\begin{equation}
z(r)^{1-\beta}-z_\texttt{max}^{1-\beta}=
z_\texttt{max}^{-1/\alpha}(r-1). \label{eq3}
\end{equation}
Combine Eq. \ref{eq1} and Eq. \ref{eq3}, we can obtain the
estimation of $z_\texttt{max}$, as
\begin{equation}
z_\texttt{max}\approx N(t)^\alpha. \label{eq4}
\end{equation}
Obviously, the text size $t$ is the sum of all words' occurrences,
say
\begin{equation}
t=\sum^{N(t)}_{r=1}z(r)\approx
\int^{N(t)}_1z(r)\texttt{d}r=\frac{z_\texttt{max}\left(N(t)^{1-\alpha}-1\right)}{1-\alpha}.
\label{eq5}
\end{equation}
Substituting $z_\texttt{max}$ by Eq. \ref{eq4}, it arrives to the
relation between $N(t)$ and $t$:
\begin{equation}
\frac{N(t)^\alpha\left(N(t)^{1-a}-1\right)}{1-\alpha}=t.
\end{equation}
The direct comparison between the empirical observation and Eq. 6,
as well as an improved version of Eq. 6, is shown in
\textbf{Materials and Methods}. Clearly, Eq. 6 is not a simply
power-law form as described by the Heaps' law. We will see that the
Heaps' law is an approximate result that can be derived from Eq. 6.
Actually, when $\alpha$ is considerably larger than 1,
$N(t)^{1-\alpha}\ll 1$ and $N(t)\approx
(\alpha-1)^{1/\alpha}t^{1/\alpha}$; while if $\alpha$ is
considerably smaller than 1, $N(t)^{1-\alpha}\gg 1$ and $N(t)\approx
(1-\alpha)t$. This approximated result can be summarized as
\begin{equation}
    \lambda=\left\{
    \begin{array}{cc}
    1/\alpha, & \alpha>1, \\
    1, & \alpha<1,
    \end{array}
    \right.
\end{equation}
which is in accordance with the previous analytical results
\cite{Cattuto2009,Baeza-Yates2000,Leijenhorst2005} for $\alpha>1$
and has complemented the case for $\alpha<1$.

Although Eq. 6 is different from a strict power law, numerical
results indicate that the relationship between $N(t)$ and $t$ can be
well fitted by the power-law functions (the fitting is usually much
better than the empirical observations about the Heaps' law, see
\textbf{Materials and Methods} for some typical examples). In Fig.
1, we report the numerical results with fixed total number of word
occurrences $t=10^5$. When $\alpha$ is considerably larger or
smaller than 1, the numerical results agree well with the known
analytical solution in Eq. 7, however, a clear deviation is observed
for $\alpha\approx 1$ (see \textbf{Materials and Methods} about how
to get the numerical results for $\alpha=1$).

To validate the numerical results of Eq. 6, we propose a stochastic
model. Given the total number of word occurrences $t$, clearly,
there are at most $t$ distinct words having the chance to appear.
The initial occurrence number of each of these $t$ words is set as
zero. At each time step, these $t$ words are sorted in descending
order of their occurrence number (words with the same number of
occurrences are randomly ordered), and the probability a word with
rank $r$ will occur in this time step is proportional to
$r^{-\alpha}$. The whole process stops after $t$ time steps. The
distribution of word occurrence always obeys the Zipf's law with a
stable exponent $\alpha$, and the growth of $N(t)$ approximately
follows the Heaps' law with $\lambda$ dependent on $\alpha$. The
simulation results about $\lambda$ vs. $\alpha$ of this model are
also reported in Fig. 1, which agree perfectly with the numerical
ones by Eq. 6. The result of the stochastic model strongly supports
the validity of Eq. 6, and thus we only discuss the numerical
results of Eq. 6.

In addition to $\alpha$, the Heaps' exponent $\lambda$ also depends
on the system size, namely the total number of word occurrences,
$t$. An example for $\alpha=1$ is shown in Fig. 2, and how $\lambda$
varies in the $(\alpha,t)$ plane is shown in Fig. 3. It is seen that
the exponent $\lambda$ increases monotonously as the increasing of
$t$. According to Eq. 6, it is obvious that in the large limit of
system size, $t\rightarrow \infty$, the exponent $\lambda$ can be
determined by the asymptotic solution Eq. 7. Actually, the
asymptotic solution well describes the systems with $\alpha\gg 1$ or
$\alpha \ll 1$ or $t\rightarrow \infty$. However, real systems are
often with $\alpha$ around 1 and of finite sizes. As indicated by
Fig. 2 and Fig. 3, the growth of $\lambda$ versus $t$ is really
slow. For example, when $\alpha=1$, for most real systems with $t$
scaling from $10^4$ to $10^8$, the exponent $\lambda$ is
considerably smaller than the asymptotic solution $\lambda=1$. Even
for very large $t$ that is probably larger than any studied real
systems, like $t=10^{16}$, the difference between numerical result
and asymptotic solution can be observed. As we will show in the next
section, this paper emphasizes the difference between empirical
observations and the asymptotic solution, and the simple numerical
method based on Eq. 6 provides a more accurate estimation.

\subsection*{Experimental Results}

We analyze a number of real systems ranging from small-scale system
containing only 40 distinct elements to large-scale system
consisting of more than $10^5$ distinct elements. The results are
listed in Table 1 while the detailed data description is provided in
\textbf{Materials and Methods}. Four classes of real systems are
considered, including the occurrences of words in different books
and different languages (data sets Nos. 1-9), the occurrences of
keywords in different journals (data sets Nos. 10-33), the confirmed
cases of the novel virus influenza A (data set No. 34), and the
citation record of PNAS articles (data set No. 35). Figure 4 reports
the Zipf's law and Heaps' law of the four typical examples, each of
which belongs to one class, respectively.

To sum up, the empirical results indicate that (i) evolving systems
displaying the Zipf's law also obey the Heaps' law even for
small-scale systems; (ii) the asymptotic solution (Eq. 7) can well
capture the relationship between the Zipf's exponent and Heaps'
exponent, and the present numerical result based on Eq. 6 can
provide considerably better estimations (the numerical results based
on Eq. 6 outperforms Eq. 7 in 34, out of 35, tested date sets).

\section*{Discussion}

Zipf's law and Heaps' law are well known in the context of complex
systems. They were discovered independently and treated as two
independent statistical laws for decades. Recently, the increasing
evidence on the coexistence of these two laws leads to serious
consideration of their relation. However, a clear picture cannot be
extracted out from the literature. For example, Montemurro and
Zanette \cite{Montemurro2002,Zanette2005} suggested that the Zipf's
law is a result from the Heaps' law while Serrano \emph{et al.}
\cite{Serrano2009} claimed that the Zipf's law can result in the
Heaps' law. In addition, many previous analyses about their relation
are based on some stochastic models, and the results are strongly
dependent on the corresponding models -- we are thus less confident
of their applicability in explaining the coexistence of the two laws
observed almost everywhere.

In this article, without the help of any specific stochastic model,
we directly show that the Heaps' law can be considered as a
derivative phenomenon given that the evolving system obeys the
Zipf's law with a stable exponent. In contrast, the Zipf's law can
not be derived from the Heaps' law without the help of a specific
model or some external conditions. However, one can not conclude
that the Zipf's law is more fundamental since there may exists some
mechanisms only resulting in the Heaps' law, namely it is possible
that a system displays the Heaps' law while does not obey the Zipf's
law. In addition, we refine the known asymptotic solution (Eq. 7) by
a more complex formula (Eq. 6), which is considerably more accurate
than the asymptotic solution, as demonstrated by both the testing
stochastic model and the extensive empirical analysis. In
particular, our investigation about the effect of system size fills
the gap in the relevant theoretical analyses.

Our analytical result (Eq. 6) indicates that the growth of
vocabulary of an evolving system cannot be exactly described by the
Heaps' law even though the system obeys a perfect Zipf's law with a
constant exponent. In fact, not only the solution of the Heaps'
exponent (Eq. 7), but also the Heaps' law itself is an asymptotic
approximation obtained by considering infinite-size systems. More
terribly, a Zipf's exponent larger than one does not correspond to a
true distribution $p(z)$ since $\langle z\rangle$ will diverge as
the increasing of the system size, yet a large fraction of real
systems can be well characterized by the Zipf's law with $\alpha>1$
(see general examples in Refs. \cite{Clauset2009,Newman2005} and
examples of degree distributions of complex networks in Refs.
\cite{Albert2002,Newman2003}). Putting the blemish in mathematical
strictness behind, the Zipf's law and Heaps' law well capture the
macroscopic statistics of many complex systems, and our analysis
provides a clear picture of their relation.

Note that, our analysis depends on an ideal assumption of a
``perfect" power law (Zipf's law) of frequency distribution, while a
real system never displays such a perfect law. Indeed, deviations
from a power law have been observed, but the assumption of a perfect
power-law distribution is widely used in many theoretical analyses.
For example, the degree distribution in email networks
\cite{Ebel2002} has a cutoff at about $z=100$ and the one in sexual
contact networks \cite{Liljeros2001} displays a drooping head, while
in the analysis of epidemic dynamics, the underlying networks are
usually supposed to be purely scale-free networks
\cite{Pastor-Satorras2001}. Another example is the study on the
effects of human dynamics on epidemic spreading
\cite{Vazquez2007,Iribarren2009}, where the interevent time
distribution of human actions are supposed as a power-law
distribution, ignoring the observed cutoffs and periodic
oscillations \cite{Zhou2008,Radicchi2009}. In a word, although the
ideal assumption of a perfect power-law distribution could not fully
reflect the reality, the corresponding analysis indeed contributes
much to our understanding of many phenomena.

An interesting implication of our results lies in the accelerated
growth of scale-free networks. Considering the degree of a node as
its occurrence frequency and the total degree of all nodes as the
text size, a growing network is analogous to a language system.
Then, the scale-free nature corresponds to the Zipf's law of word
frequency and the accelerated growth corresponds to the Heaps' law
of the vocabulary growth. In an accelerated growing network, the
total degree $t$ (proportional to the number of edges) scales in a
power-law form as $t\sim N(t)^\phi$, where $N(t)$ denotes the number
of nodes and $\phi>1$ is the accelerating exponent. At the same
time, the degree distribution usually follows a power law as
$p(k)\sim k^{-\beta}$ where $k$ denotes the node degree. For
example, the Internet at the autonomous system (AS) level displays
the scale-free nature with $\beta\approx 2.16$ (see Table 1 in Ref.
\cite{Caldarelli2007}) and thus $\alpha=\frac{1}{\beta-1}\approx
0.862$. According to a recent report \cite{Zhang2008b} on empirical
analysis of the Internet at the AS level, till December 2006, the
total degree is $t=105652$. The corresponding numerical result of
the Heaps' exponent is $\lambda\approx 0.92$ and thus the
accelerating exponent can be estimated as
$\phi=\frac{1}{\lambda}\approx 1.09$. In contrast, the asymptotic
solution Eq. 7 suggests a steady growing as $\phi=\lambda=1$.
Compared with the empirical result $\phi\approx 1.11$
\cite{Zhang2008b}, Eq.~6 ($\phi=1.09$) gives better result than Eq.
7 ($\phi=1$). Actually, the asymptotic solution indicates that all
the scale-free networks with $\beta>2$ should grow in a steady
(linear) manner, which is against many known empirical observations
\cite{Dorogovtsev2001,Smith2007,Broder2000,Zhang2008b}, while the
refined result in this article is in accordance with them. Further
more, our result provides some insights on the growth of complex
networks, namely the accelerated growth can be expected if the
network is scale-free with a stable exponent and this phenomenon is
prominent when $\beta$ is around 2.

\section*{Materials and Methods}

\subsection{Relation between Zipf's Law and Power Law}

Given the Zipf's law $z(r)\sim r^{-\alpha}$, we here prove that the
probability density function $p(z)$ obeys a power law as $p(z)\sim
z^{-\beta}$ with $\beta=1+\frac{1}{\alpha}$. Considering the data
points with ranks between $r$ and $r+\delta r$ where $\delta r$ is a
very small value. Clearly, the number of data points is $\delta r$,
which can be expressed by the probability density function as
\begin{equation}
\delta r = p(z(r))\delta z,
\end{equation}
where
\begin{equation}
\delta z\sim r^{-\alpha}-(r+\delta r)^{-\alpha}\sim
r^{-\alpha-1}\delta r.
\end{equation}
Therefore, we have
\begin{equation}
p(r^{-\alpha})\sim r^{-\alpha-1}\sim
(r^{-\alpha})^{-\frac{\alpha+1}{\alpha}},
\end{equation}
namely $\beta=1+\frac{1}{\alpha}$. Analogously, the Zipf's law
$z(r)\sim r^{-\alpha}$ can be derived from the power-law probability
density distribution $p(z)\sim z^{-\beta}$, with
$\alpha=\frac{1}{\beta-1}$.

\subsection{Direct Comparison between Empirical and Analytical Results}

Given the parameter $\alpha$, according to Eq. 6, we can numerically
obtain the function $N(t)$. The comparison between Eq. 6 and the
empirical data for words in the book ``La Divina Commedia" and
keywords in the PNAS articles are shown in Fig. \ref{Direct
Comparsion}. The growing tendency of distinct words can be well
captured by Eq. 6. Actually, using a more accurate normalization
condition $\int^{z_\texttt{max}+\frac{1}{2}}_\frac{1}{2}
p(z)\texttt{d}z=1$, as an improved version of Eq. 4, the estimation
of $z_\texttt{max}$ is determined by
\begin{equation}
N(t)z_\texttt{max}^{-\frac{1}{\alpha}}=\left(\frac{1}{2}\right)^{-\frac{1}{\alpha}}-\left(z_\texttt{max}+\frac{1}{2}\right)^{-\frac{1}{\alpha}}.
\label{improved analysis}
\end{equation}
Given the parameter $\alpha$, for an arbitrary $N(t)$, one can
estimates the corresponding $z_\texttt{max}$ according to Eq.
\ref{improved analysis} and then determines the value of $t$ by Eq.
5. The numerical results of this improved version are also presented
in Fig. \ref{Direct Comparsion}, which fits better than Eq. 6 to the
empirical data. Notice that, both the two analytical results give
almost the same slope in the log-log plot of $N(t)$ function, namely
the Heaps' exponents obtained by these two versions are almost the
same.

\subsection{Examples of Numerical Results}

Mathematically speaking, as indicated by Eq. 6, $N(t)$ does not
scale in a power law with $t$. However, the numerical results
suggest that the dependence of $N(t)$ on $t$ can be well
approximated as power-law functions. As shown in Fig.
\ref{examples}, for a wide range of $\alpha$, $N(t)$ can be well
fitted by $t^\lambda$, and the value of fitting exponent $\lambda$
depends on both $\alpha$ and $t$.

\subsection{The case of $\alpha=1$}

The numerical solution of Eq. 6 for $\alpha=1$ can be obtained by
considering the limitation $\alpha \rightarrow 1$, where
$N(t)^{\alpha}\approx N(t)$ and $N(t)^{1-\alpha}\approx
1+(1-\alpha)\ln N(t)$. Accordingly, Eq. 6 can be rewritten as
\begin{equation}
N(t)\ln N(t)=t. \label{a=1}
\end{equation}
When $t$ approaches to infinity, $N(t)$ scales almost linearly with
$t$ since $\lim_{t\rightarrow \infty}\frac{\ln N(t)}{N(t)}=0$.
Actually, the solution can be expressed as $N(t)=t/W(t)$ where
$W(t)$ is the well-known \emph{Lambert W function}
\cite{Corless1996} that satisfies
\begin{equation}
W(t)e^{W(t)}=t.
\end{equation}
For any finite system, the numerical result can be produced by Eq.
\ref{a=1}.

\subsection{Data description}

The data sets analyzed in this article can be divided into four
classes. According to the data sets shown in Table 1, these four
classes are as follows.

(i) Occurrences of words in different books and different languages
(data sets Nos. 1-9). The data set No. 1 is the English book
(\emph{Moby Dick}) written by Herman Melville; the data sets No. 2
(\emph{De Bello Gallico}), No. 3 (\emph{Philosophi{\ae} Naturalis
Principia Mathematica}) and No. 7 (\emph{Aeneis}) are Latin books
written by Gaius Julius Caesar, Isaac Newton and Virgil
respectively; the data sets No. 4 (\emph{Don Quijote}), No. 5
(\emph{La Celestina}) and No. 8 (\emph{Cien a\"nos de soledad}) are
Spanish novels written by Miguel de Cervantes, Fernando de Rojas and
Gabriel Garc\'ia M\'arquez, respectively; the data set No. 6
(\emph{Faust}) is a German opera written by Johann Wolfgang von
Goethe; the data set No. 9 (\emph{La Divina Commedia di Dante}) is
the Italian epic poem written by Dante Alighieri. All the above data
are collected by Carpena \emph{et al.} \cite{Carpena2009} and
available at http://bioinfo2.ugr.es/TextKeywords/index.html.

(ii) Occurrences of keywords in different journals (data sets Nos.
10-33). These 24 journals, from No. 10 to No. 33 are PNAS, Chin.
Sci. Bull., J. Am. Chem. Soc., Acta Chim. Sinica, Crit. Rev.
Biochem. Mol. Biol., J. Biochem., J. Nutr. Biochem., Phys. Rev.
Lett., Appl. Phys. Lett., Physica A, ACM Comput. Surv., ACM Trans.
Graph., Comput. Netw., ACM Trans. Comput. Syst., Econmetrica, J.
Econ. Theo., SIAM Rev., SIAM J. Appl. Math., Invent. Math., Ann.
Neurol., J. Evol. Biol., Theo. Popul. Biol., MIS Quart., and IEEE
Trans. Automat. Contr.. These data are collected from the ISI Web of
Knowledge (http://isiknowledge.com/). For every scientific journal,
we consider the keywords sequence in each article according to its
publishing time. Since most of the published articles do dot have
keywords before 1990 in ISI database, we limit our collections from
1991 to 2007 (except for ACM Comput. Surv. which is available only
from 1994 to 1999).

(iii) Confirmed cases of the novel virus influenza A (data set No.
34). The data of the cumulative number of laboratory confirmed cases
of H1N1 of each country are available from the website of Epidemic
and Pandemic Alert of World Health Organization (WHO)
(http://www.who.int/). The analyzed data set reported influenza A
starting from April 26 to May 18, updated each one or two days.
After May 18, the distribution of confirmed cases in each country
shifted from a power law to a power-law form with exponential cutoff
\cite{Han2009}.

(iv) Citation record of PNAS articles (data set No. 35). This data
set consists of all the citations to PNAS articles from papers
published between 1915 and 2009 according to the ISI database,
ordered by time.




\newpage

\begin{figure}
\centering
\includegraphics[width=0.8\textwidth]{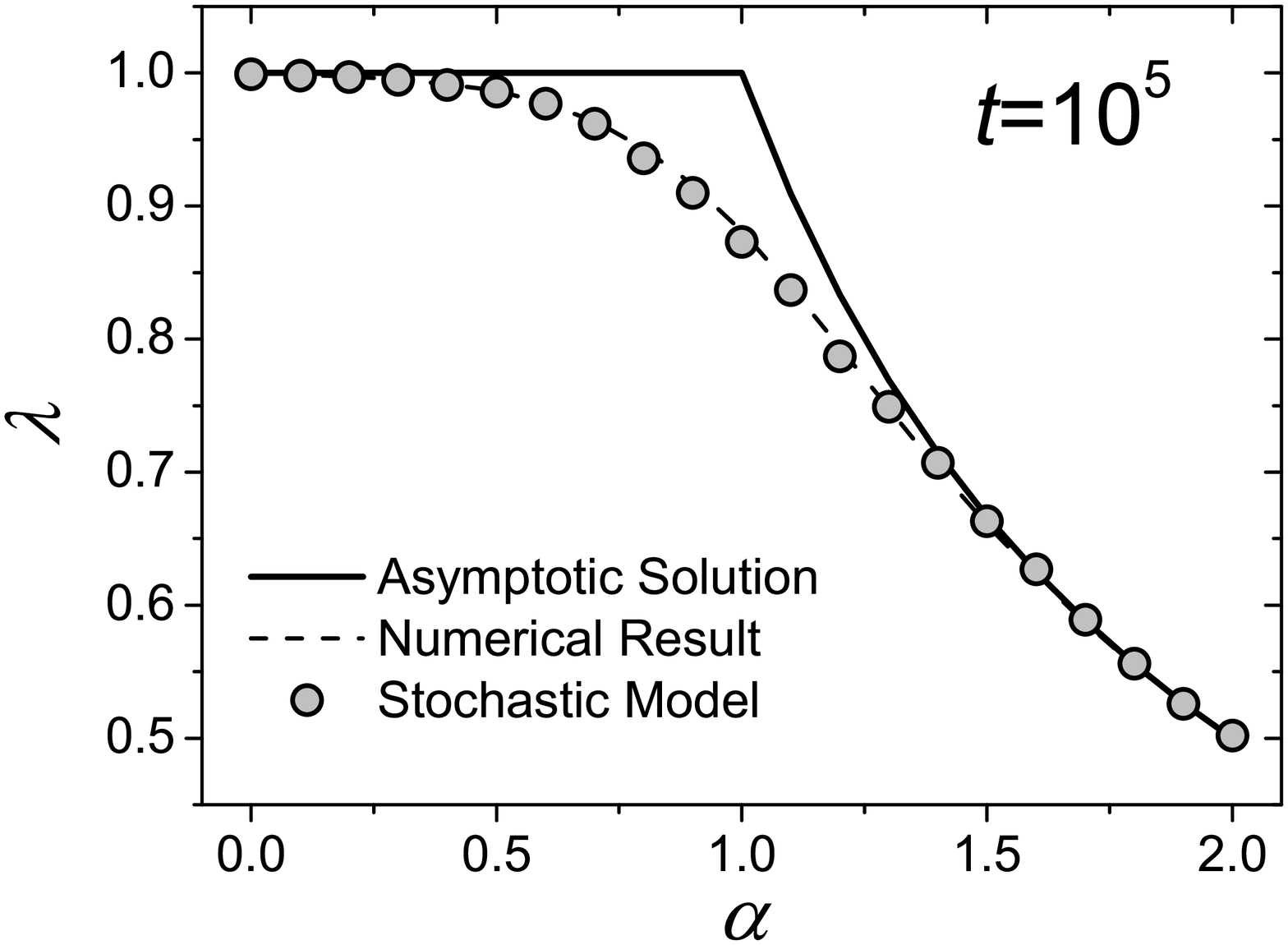}
\caption{\textbf{Relationship between the Heaps' exponent $\lambda$
and Zipf's exponent $\alpha$.} The solid curve represents the
asymptotic solution shown in Eq. 7, the dash curve is the numerical
result based on Eq. 6, and the circles denote the result from the
stochastic model. For the numerical result and the result of the
stochastic model, the total number of word occurrences is fixed as
$t=10^5$.}
\end{figure}

\newpage

\begin{figure}
\centering
\includegraphics[width=0.8\textwidth]{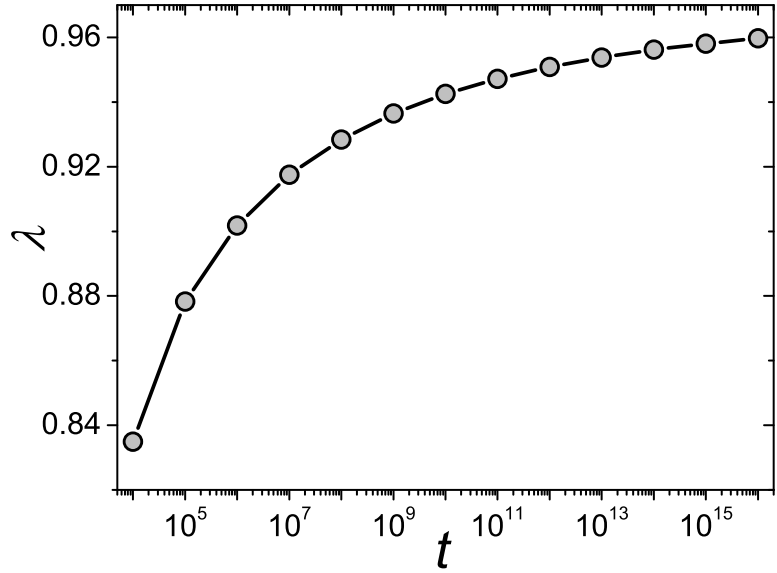}
\caption{\textbf{Effect of system size on the Heaps' exponent
$\lambda$.} The Zipf's exponent is fixed as $\alpha=1$.}
\end{figure}

\newpage
\begin{figure}
\centering
\includegraphics[width=0.8\textwidth]{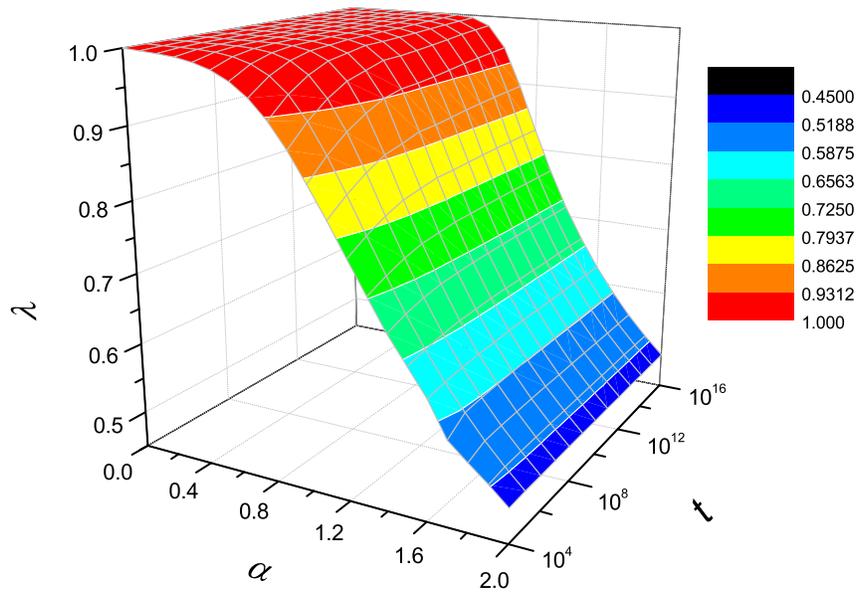}
\caption{\textbf{Heaps' exponent $\lambda$ as a function of
$(\alpha,t)$}.}
\end{figure}

\newpage

\begin{figure}
\centering
\includegraphics[width=0.4\textwidth]{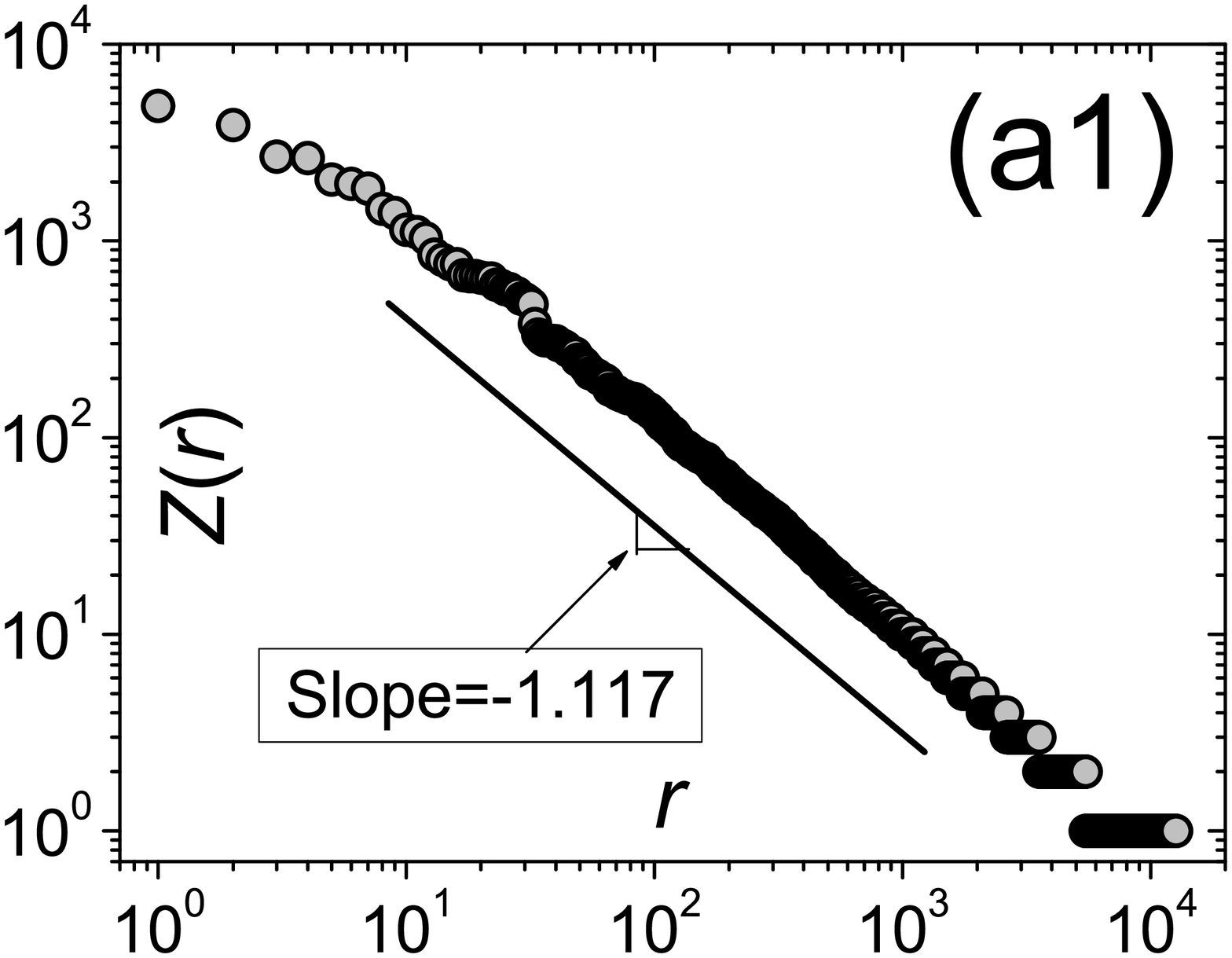}
\includegraphics[width=0.4\textwidth]{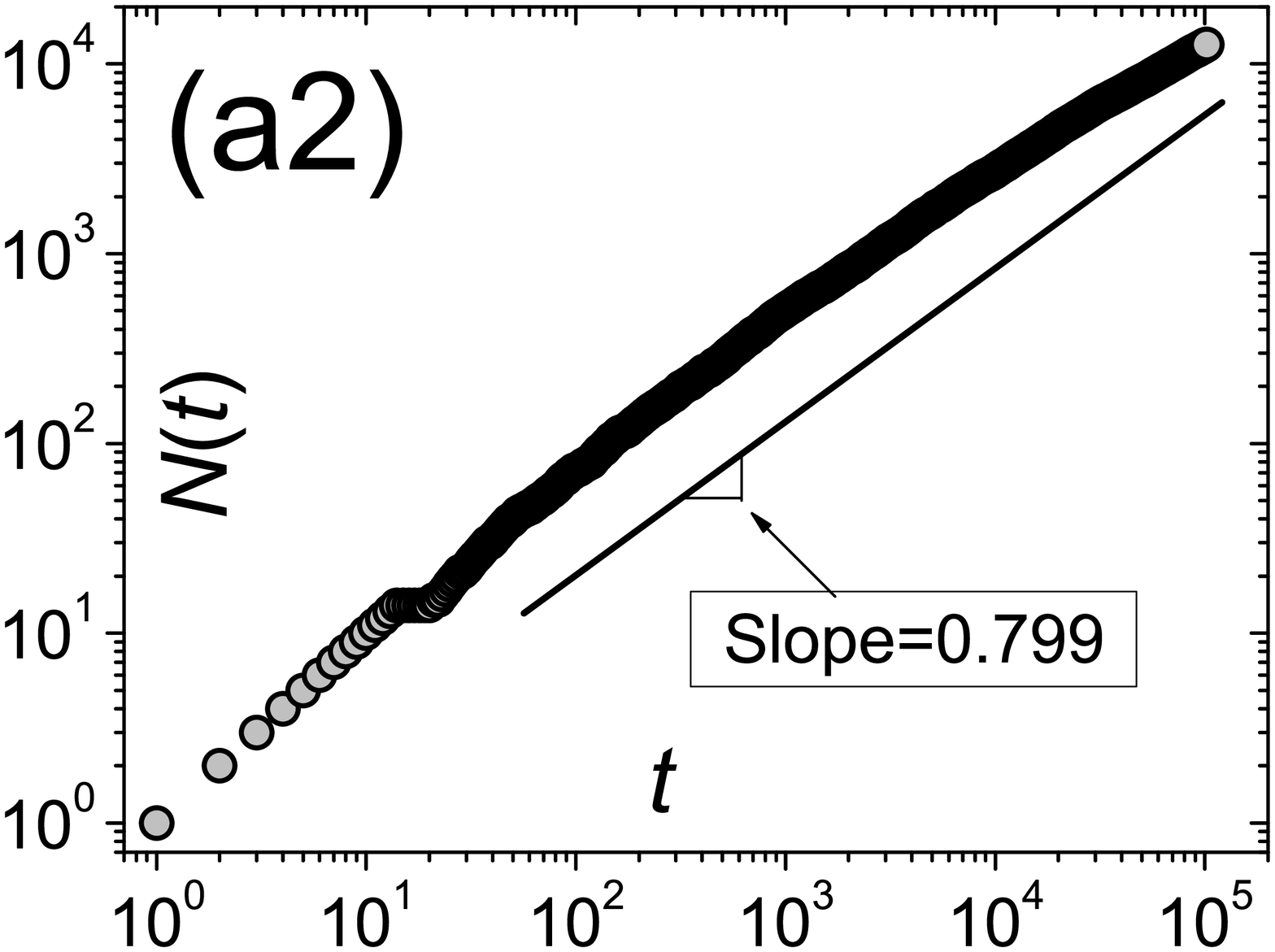}
\includegraphics[width=0.4\textwidth]{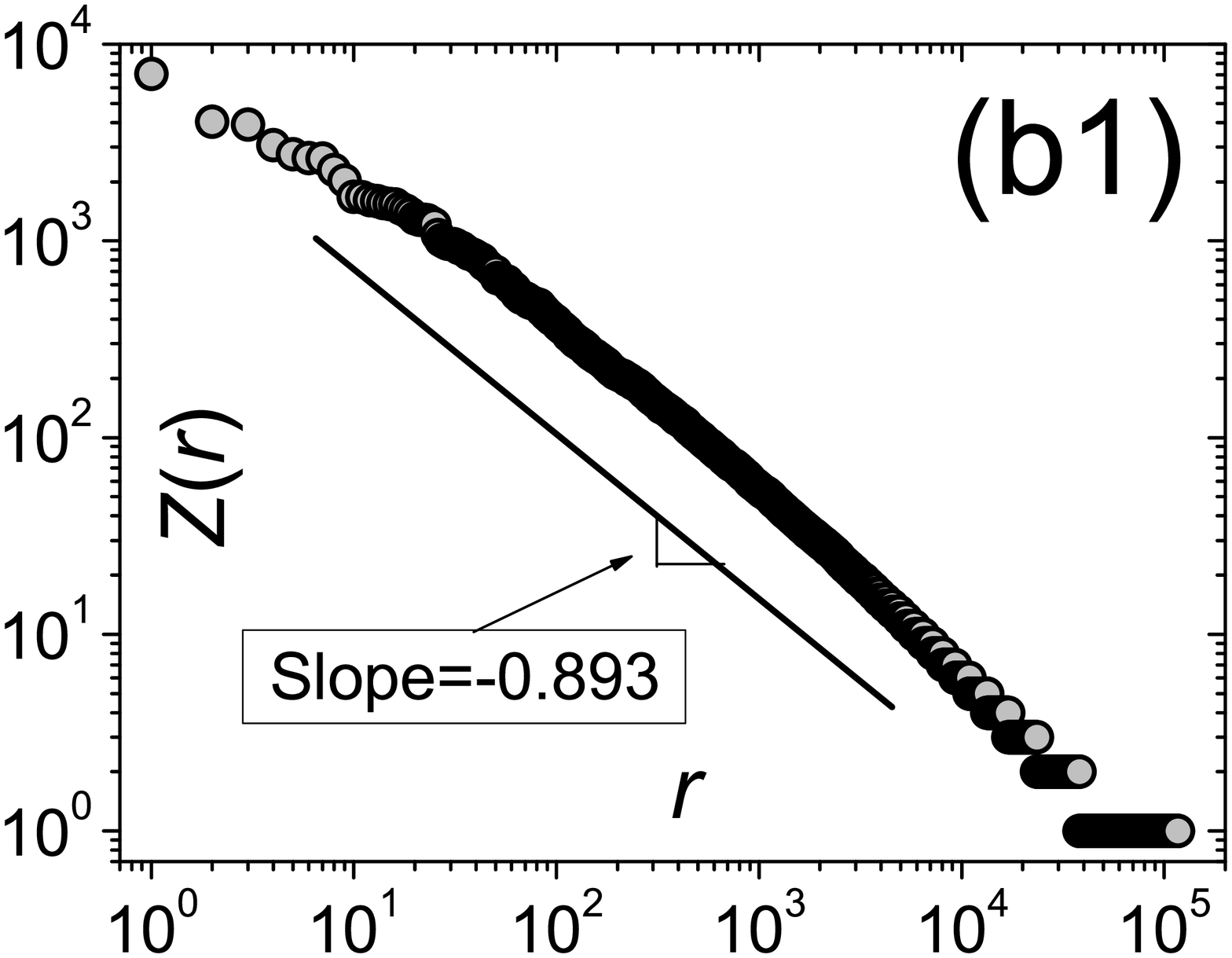}
\includegraphics[width=0.4\textwidth]{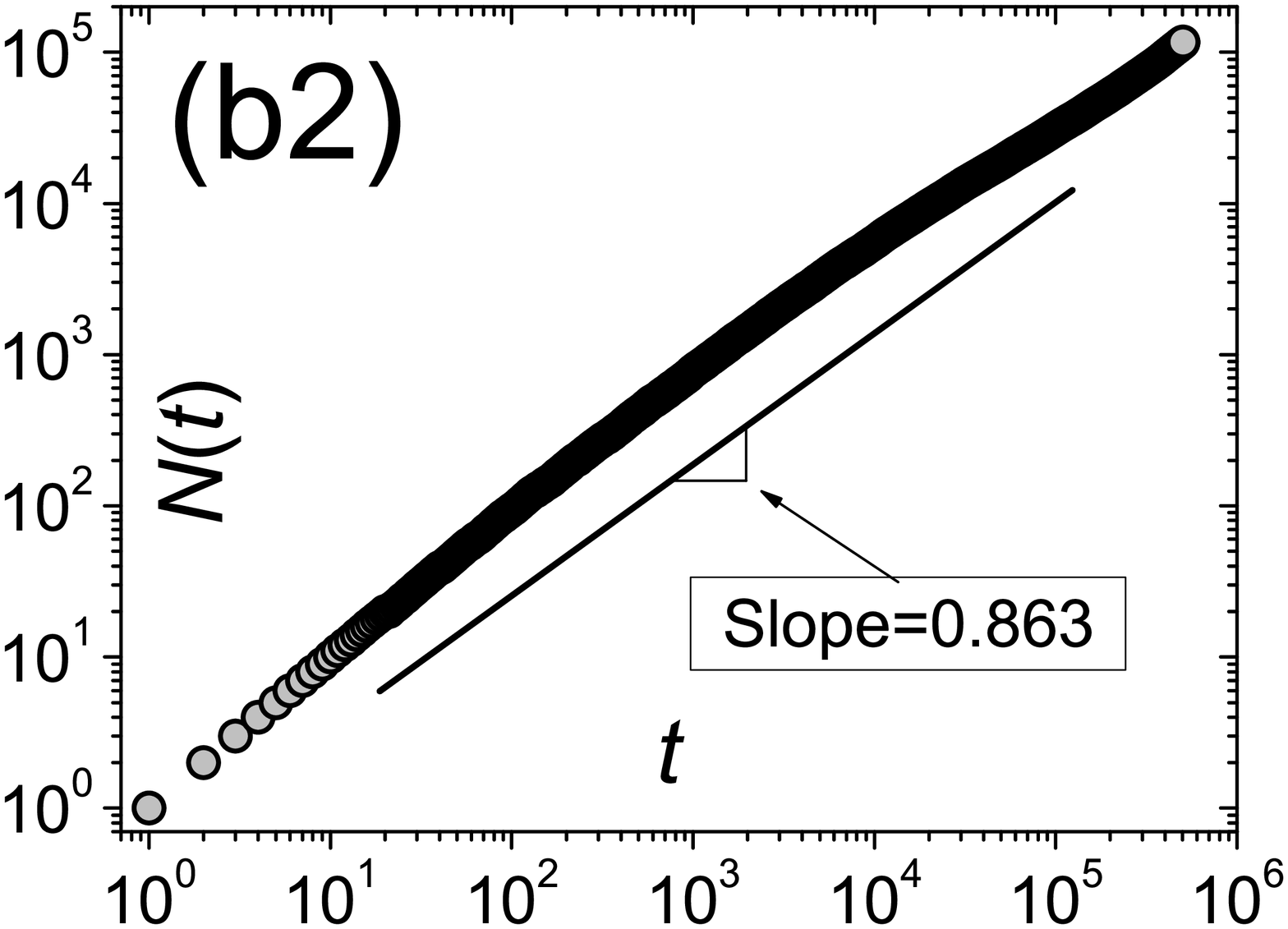}
\includegraphics[width=0.4\textwidth]{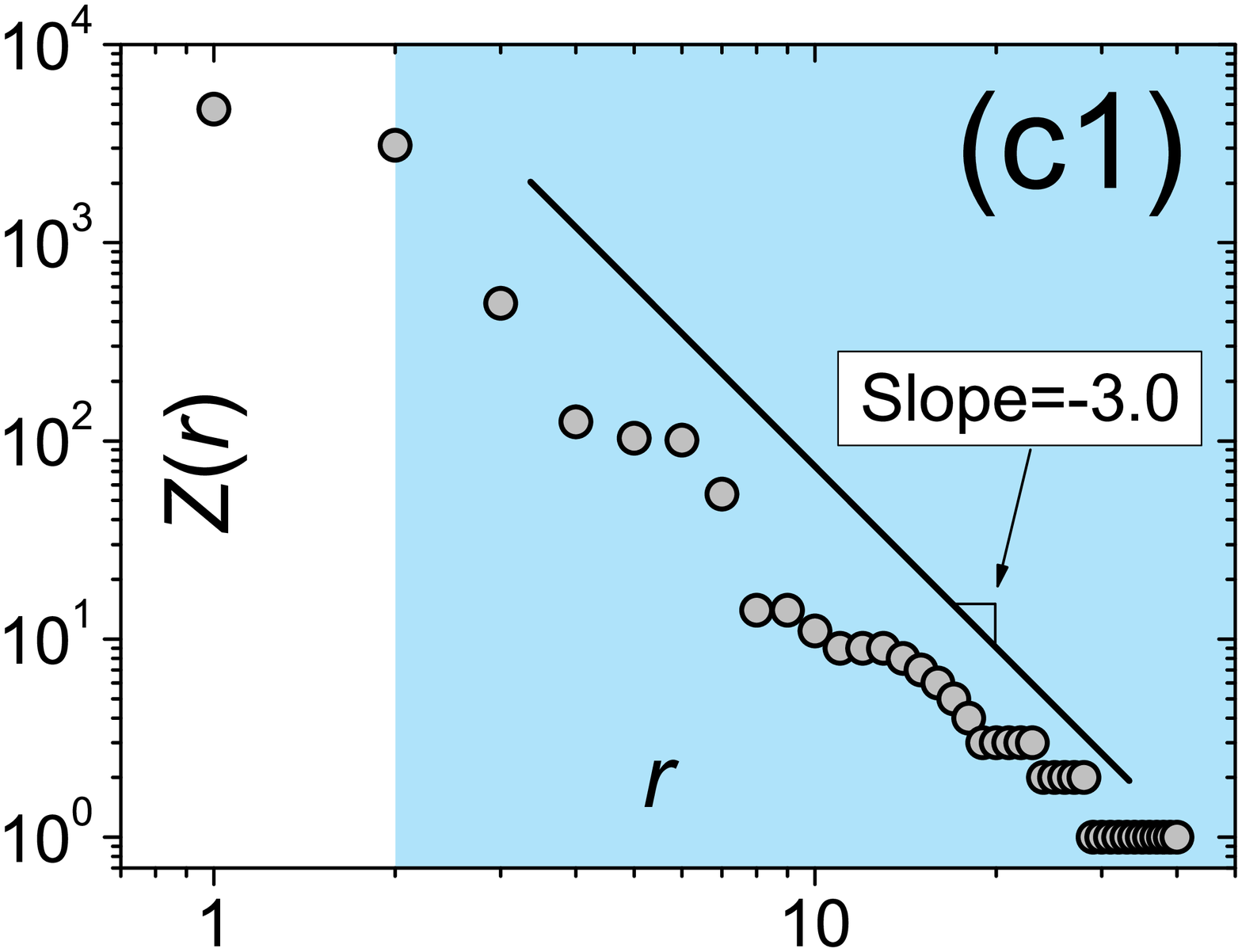}
\includegraphics[width=0.4\textwidth]{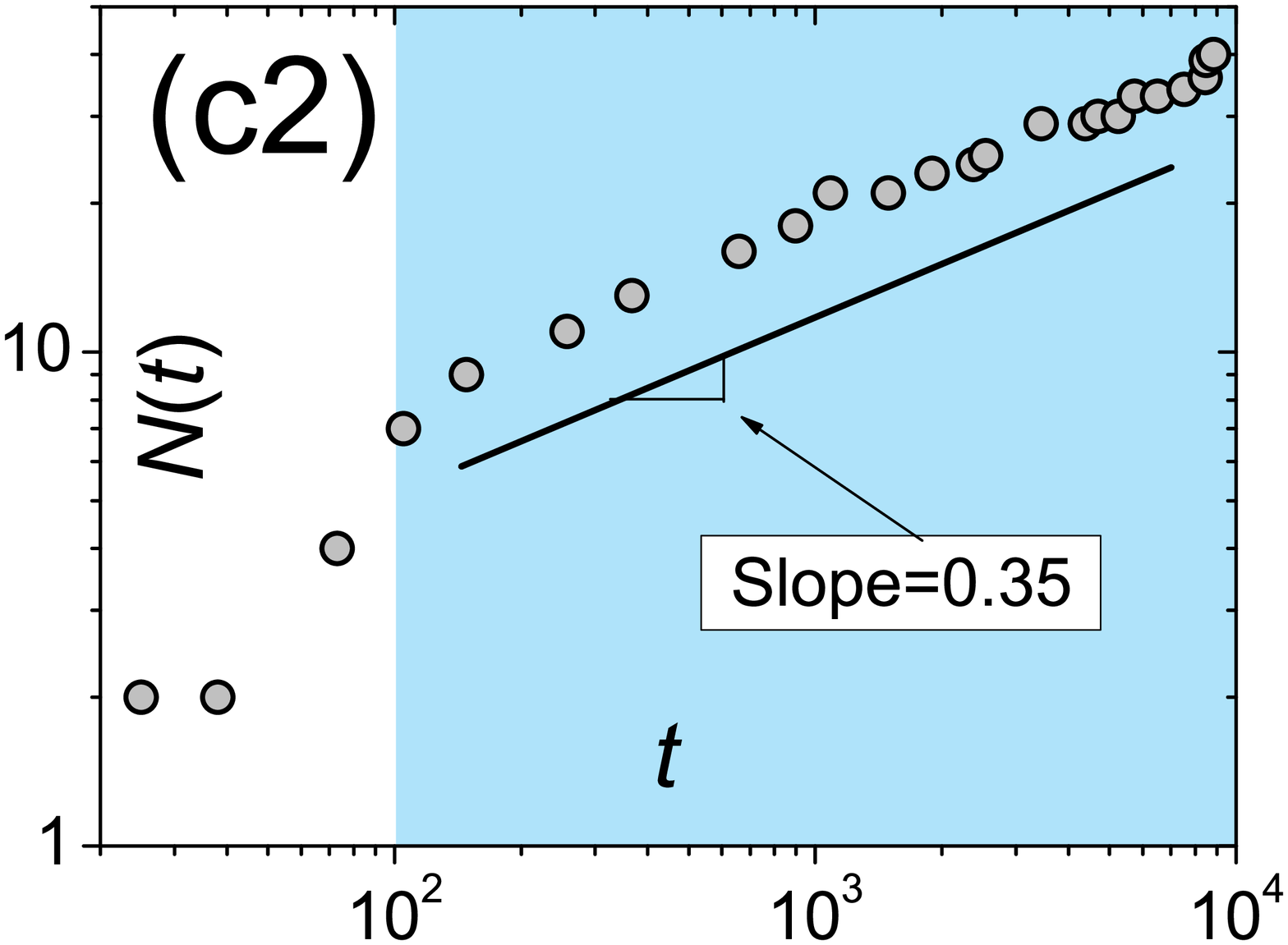}
\includegraphics[width=0.4\textwidth]{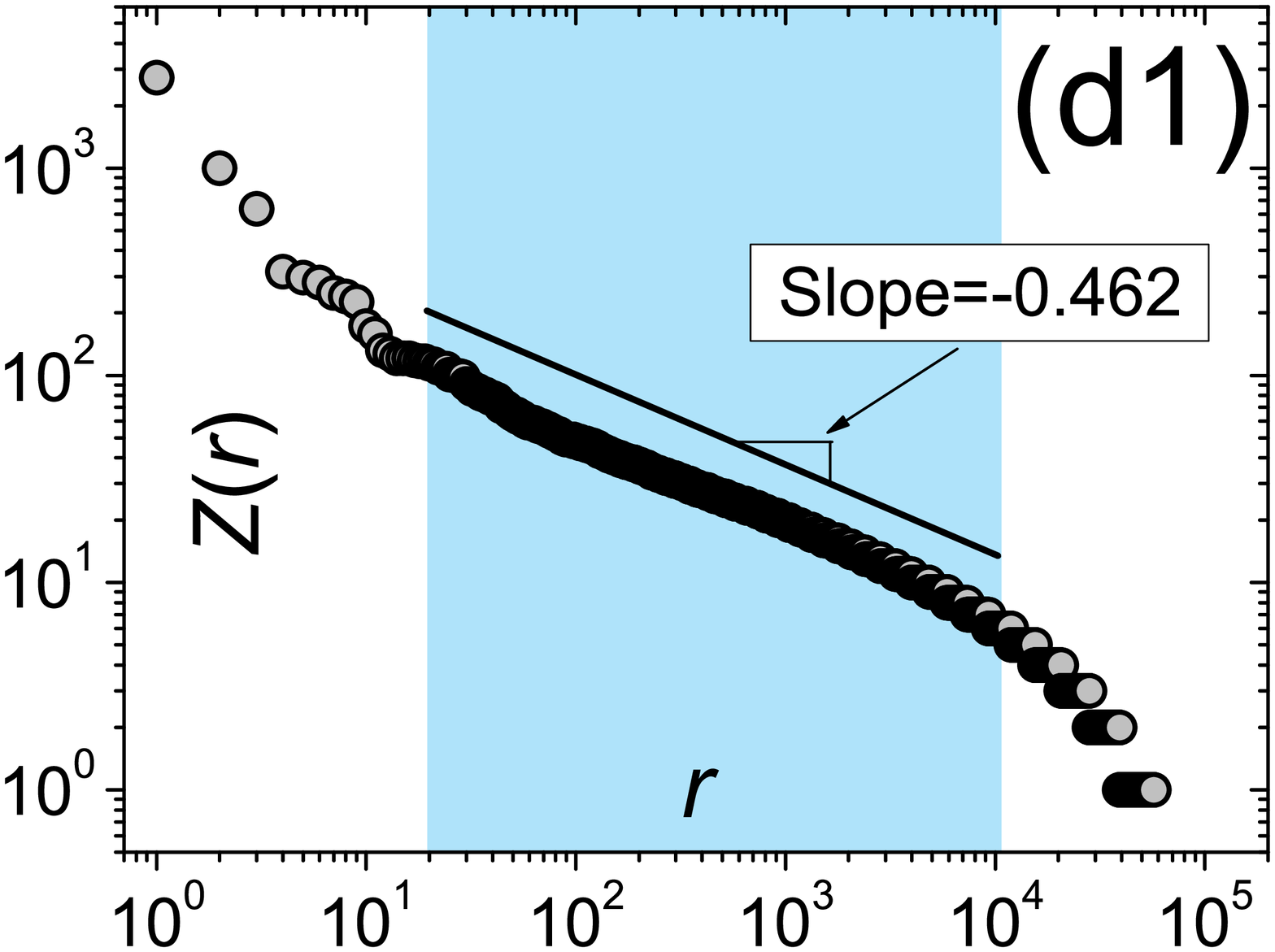}
\includegraphics[width=0.4\textwidth]{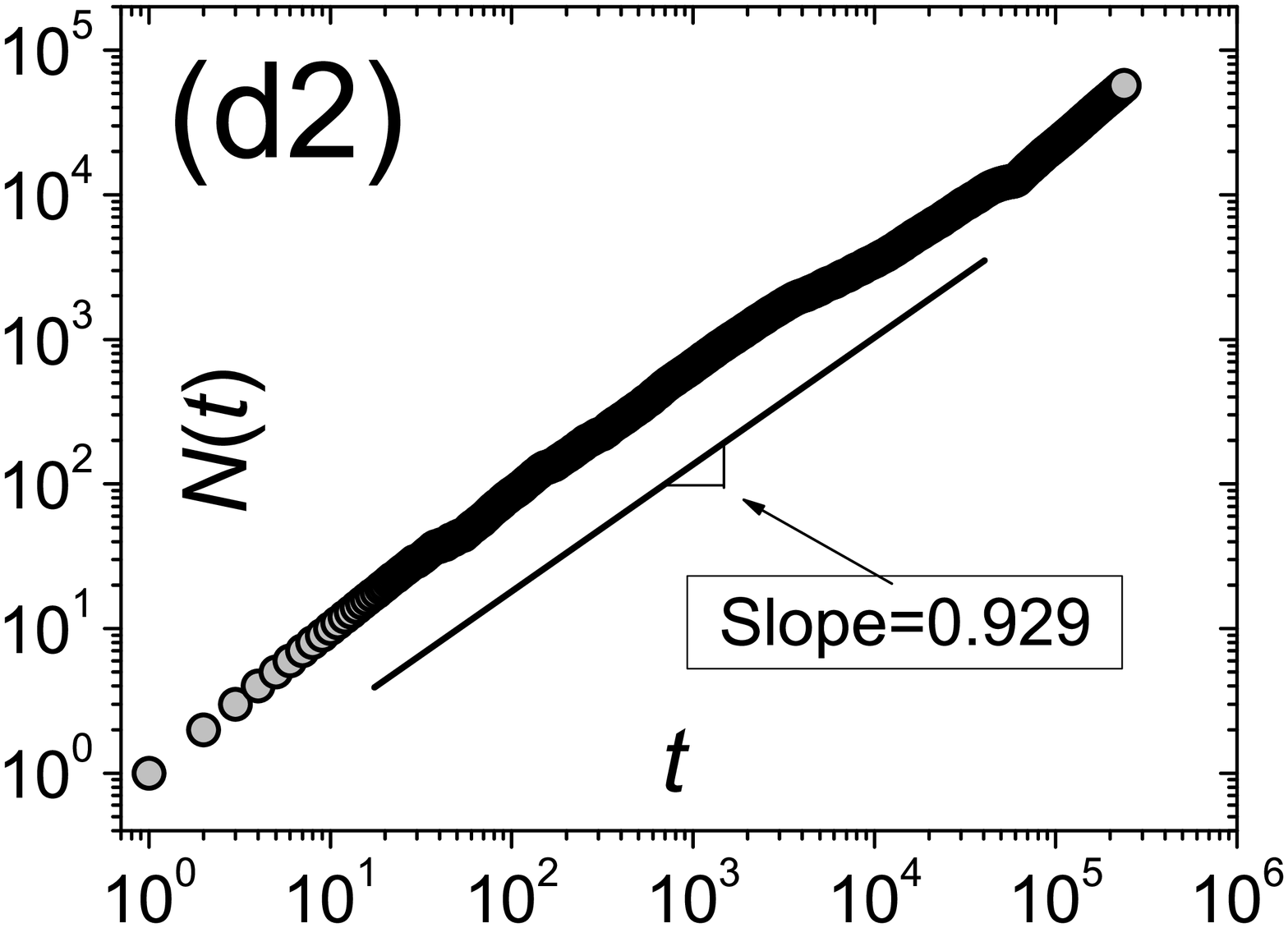}
\caption{\textbf{Zipf's law and Heaps' law in four example systems}.
(a) Words in Dante Alghieri's great book ``La Divina Commedia" in
Italian \cite{Carpena2009} where $Z(r)$ is the frequency of the word
ranked $r$ and $N(t)$ is the number of distinct words. (b) Keywords
of articles published in the Proceedings of the National Academy of
Sciences of the United States of America (PNAS) \cite{Zhang2008}
where $Z(r)$ is the frequency of the keyword ranked $r$ and $N(t)$
is the number of distinct keywords; (c) Confirmed cases of the novel
virus influenza A (H1N1) \cite{Han2009} where $Z(r)$ is the number
of confirmed cases of the country ranked $r$ and $N(t)$ is the
number of infected country in the presence of $t$ confirmed cases
over the world; (d) PNAS articles having been cited at least once
from 1915 to 2009 where $Z(r)$ is the number of citations of the
article ranked $r$ and $N(t)$ is the number of distinct articles in
the presence of $t$ citations to PNAS. In (c), the data set is small
and thus the effective number is only two digits. The fittings in
(c1) and (c2) only cover the area marked by blue. In (d1), the
deviation from a power law is observed in the head and tail, and
thus the fitting only covers the blue area. The Zipf's (power-law)
exponents and Heaps' exponents are obtained by using the
\emph{maximum likelihood estimation}
\cite{Clauset2009,Goldstein2004} and \emph{least square method},
respectively. Statistics of these data sets can be found in Table 1
(the data set numbers of (a), (b), (c) and (d) are 9, 10, 34 and 35
in Table 1) with detailed description in \textbf{Materials and
Methods}.}
\end{figure}

\newpage

\begin{figure}
\centering
\includegraphics[width=0.48\textwidth]{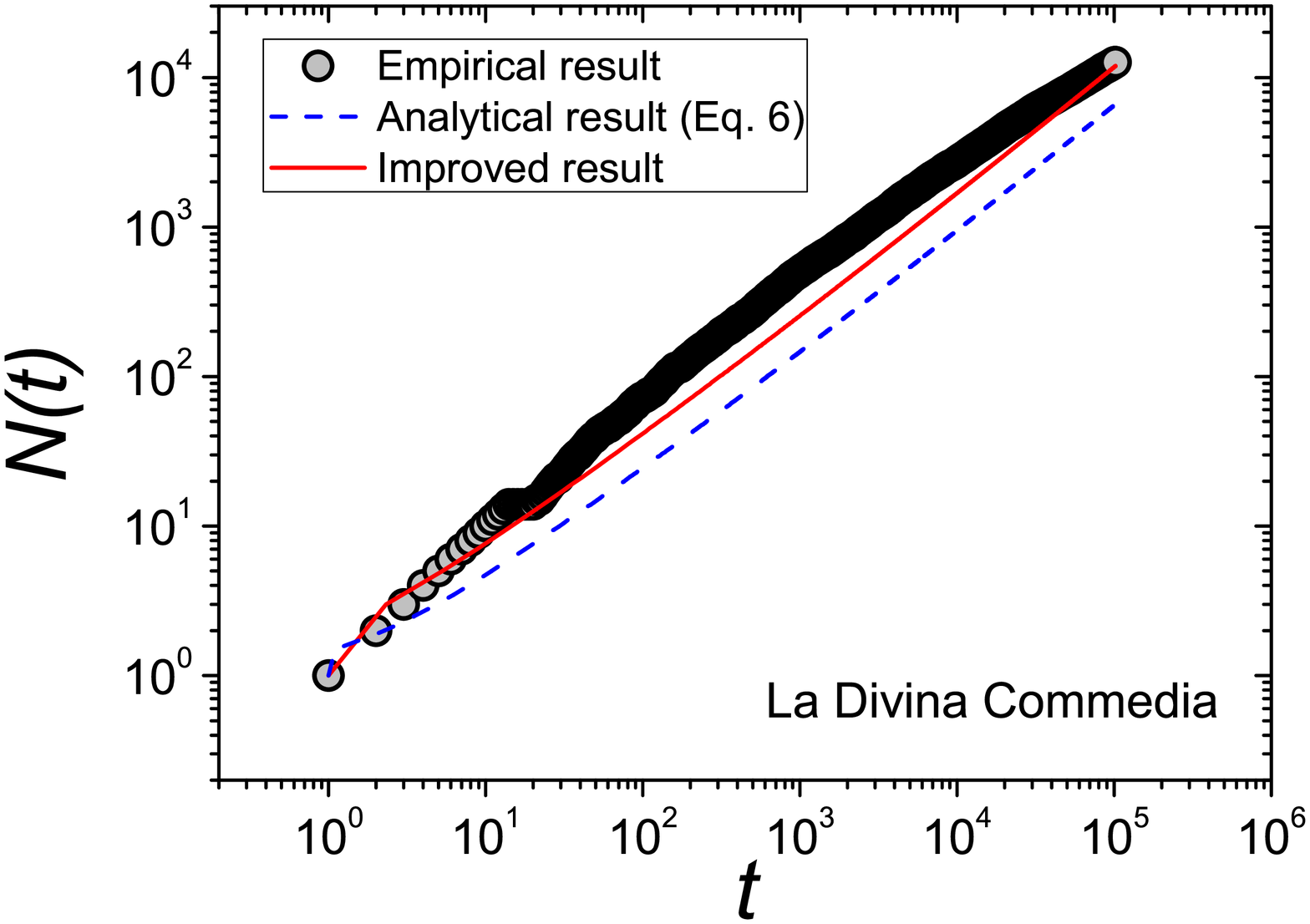}
\includegraphics[width=0.48\textwidth]{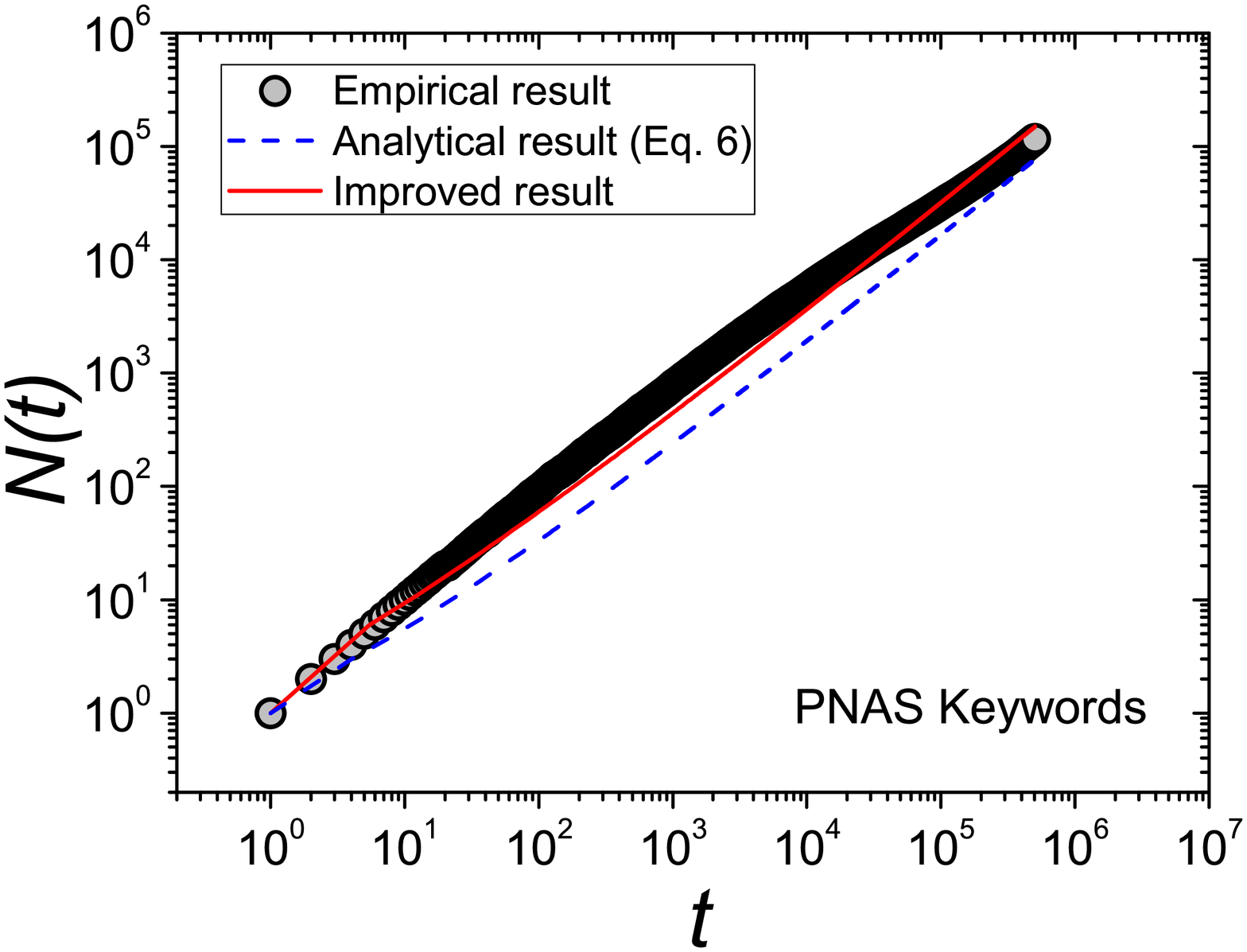}
\caption{\textbf{Direct comparison between the empirical data and
Eq. 6 as well as its improved version.} The left and right plots are
for the words in ``La Divina Commedia" and the keywords in PNAS. The
blue dash lines and red solid lines present the results of Eq. 6 and
Eq. \ref{improved analysis}, respectively. In accordance with Figure
4 and Table 1, the values of the parameter $\alpha$ are given as
1.117 and 0.893, respectively.} \label{Direct Comparsion}
\end{figure}

\newpage

\begin{figure}
\centering
\includegraphics[width=0.32\textwidth]{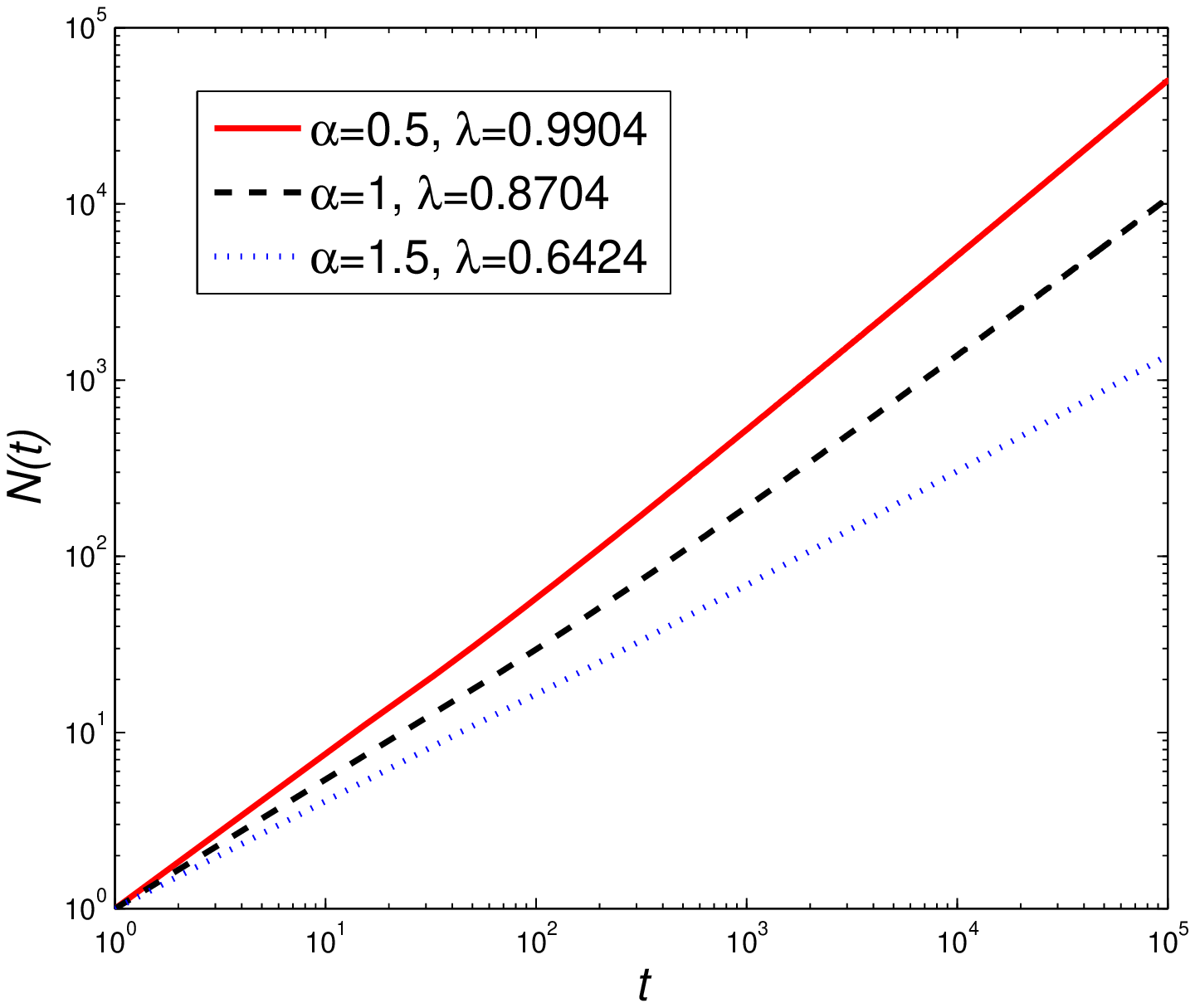}
\includegraphics[width=0.32\textwidth]{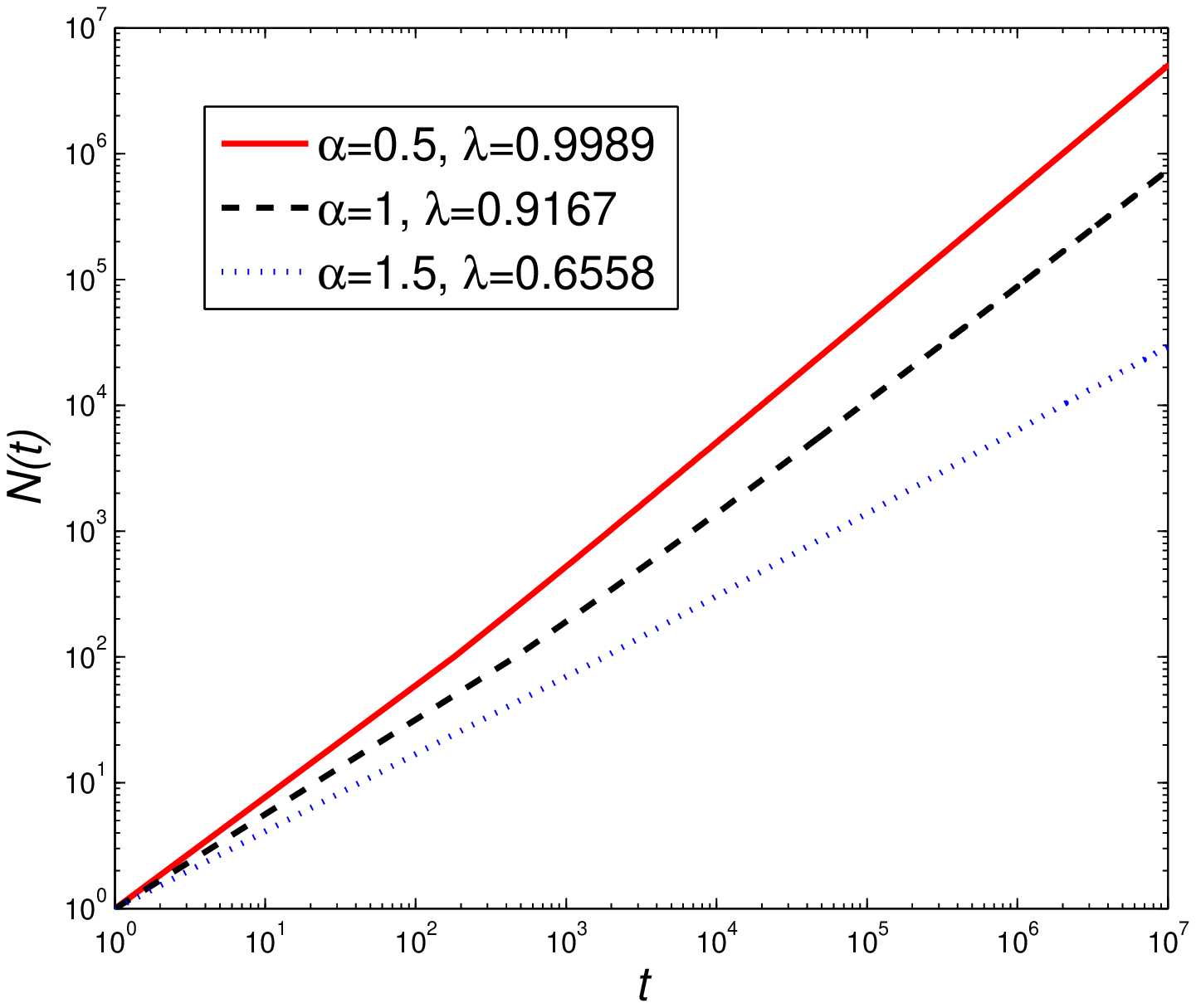}
\includegraphics[width=0.32\textwidth]{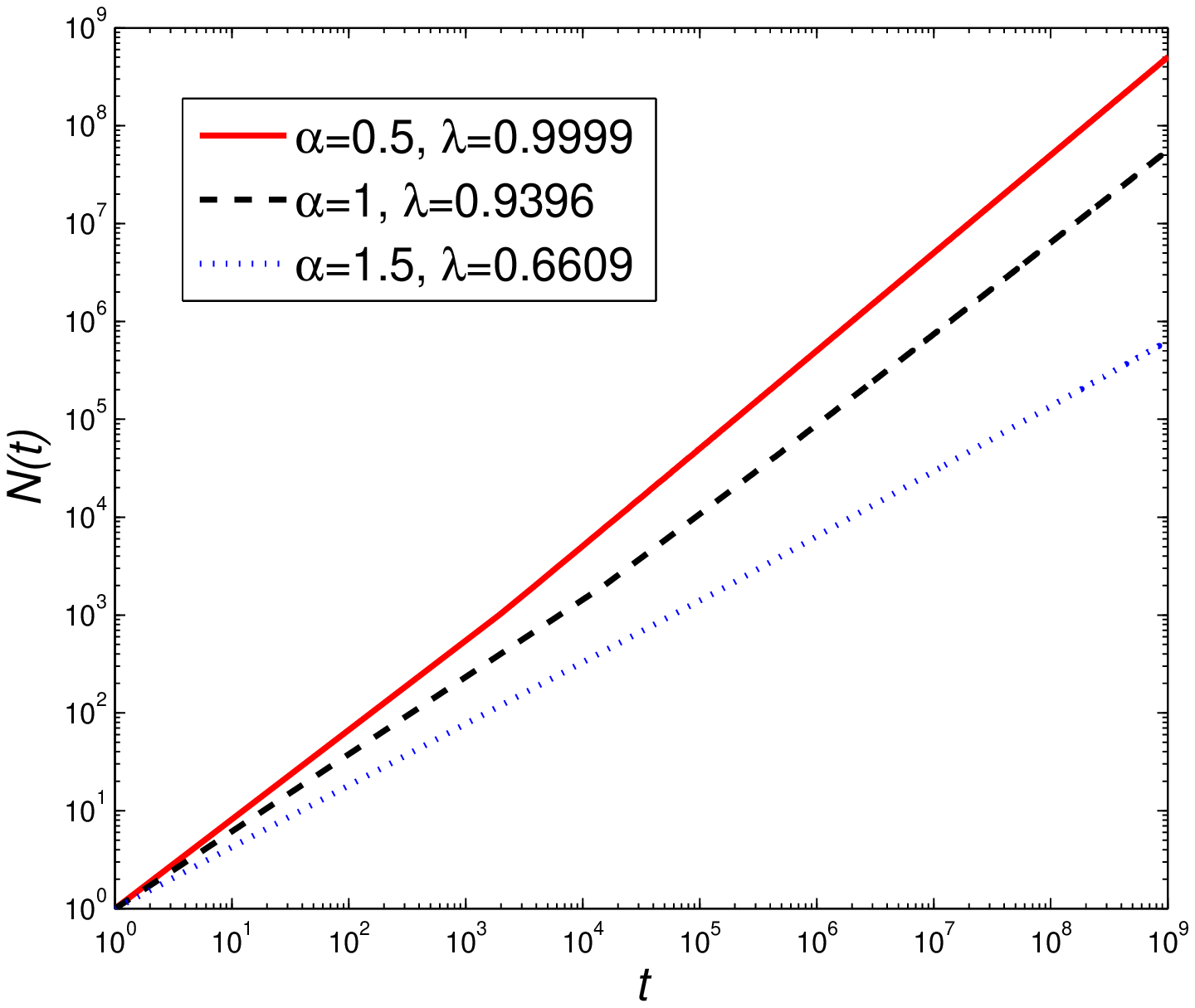}
\caption{\textbf{$N(t)$ vs. $t$ according to the numerical results
of Eq. 6.} The red, black and blue line corresponding to the cases
of $\alpha=0.5$, $\alpha=1.0$ and $\alpha=1.5$. The system sizes
(i.e., the total number of word occurrences), from left to right,
are $t=10^5$, $t=10^7$ and $10^9$. Fitting exponent $\lambda$ is
obtained by the \emph{least square method}. The fitting lines and
numerical results almost completely overlap.} \label{examples}
\end{figure}

\newpage

\begin{table}
\centering \caption{\textbf{Empirical statistics and analysis
results of real data sets}. $T$ is the total number of elements,
$N(T)$ is the total number of distinct elements, $\alpha$ is the
Zipf's exponent obtained by the \emph{maximum likelihood estimation}
\cite{Clauset2009,Goldstein2004}, $\lambda_a$ is the asymptotic
solution of the Heaps' exponent as shown in Eq. 7, $\lambda_n$ is
the numerical value of the Heaps' exponent given $T$ and $\alpha$ as
shown in Fig. 3, and $\lambda_e$ is the empirical result of the
Heaps' exponent obtained by the \emph{least square method}. The
effective number of the 34th data set is only two digits since the
size of this data set is very small. Except the 4th data set, in all
other 34 real data sets, the numerical results based on Eq. 6
outperform the asymptotic solution shown in Eq. 7. Detailed
description of these data sets can be found in \textbf{Materials and
Methods}.}
\begin{tabular}{ccccccc}
\hline \hline No. & $T$ & $N(T)$ & $\alpha$ & $\lambda_{a}$ &
$\lambda_{n}$ & $\lambda_{e}$ \\
\hline 1 & 206779 & 18217& 1.323&0.756 & 0.725 & 0.738\\ \hline 2
&20516& 5671& 0.969& 1 & 0.858 & 0.859 \\\hline 3 &109854 &13906
&1.063 & 0.941& 0.845 &0.817\\\hline 4 & 449205& 20220&1.464 &0.683
&0.667 & 0.679 \\\hline 5 &68458 &9191 &1.095 &0.913 &0.823 & 0.810
\\\hline 6 & 81037& 13254& 1.025& 0.976& 0.859& 0.832 \\\hline 7 &
63742&16622 &1.057 & 0.946&0.840 &0.852 \\\hline 8 &138985 &15550 &
1.188& 0.842&0.787 & 0.765 \\\hline 9 &101940 &12667 & 1.117&
0.895&0.818 &0.799 \\\hline 10 & 504610& 116800& 0.893& 1& 0.936&
0.863 \\\hline 11 &53214 &34194 &0.540 &1 &0.983 & 0.946 \\\hline 12
&310853 &69185 &0.939 &1 &0.913 &0.871 \\\hline 13 &30852 &17562
&0.595 &1 &0.972 &0.939 \\\hline 14 &2761 &2328 &0.397 & 1& 0.964&
0.978 \\\hline 15 &58300 &22599 & 0.786& 1& 0.941& 0.914 \\\hline 16
& 20660&8155 & 0.790& 1&0.921 &0.890 \\\hline 17 & 226090& 69251&
0.692&1 &0.977 &0.894 \\\hline 18&176291 &62567 &0.572 &1 &0.989 &
0.920 \\\hline 19 & 44735& 19933&0.685 &1 & 0.961 &0.915 \\\hline 20
&1924 & 1323& 0.463& 1& 0.946& 0.939 \\\hline 21 & 5093& 2985&0.593
&1&0.941 &0.920 \\\hline 22 &3490 &2442 &0.500 &1 &0.952 &0.950
\\\hline 23 & 1403& 787& 0.524& 1& 0.926&0.931 \\\hline 24 &7469
&4142 & 0.654& 1& 0.936&0.925 \\\hline 25 & 7710& 3857& 0.658& 1&
0.935&0.930 \\\hline 26 &3232 &2658 &0.416 &1&0.964 &0.976 \\\hline
27 &13165 &7743 &0.612 & 1& 0.959&0.936 \\\hline 28 &3749 &2353
&0.568 &1 &0.943 & 0.940 \\\hline 29 &30092& 11002& 0.815& 1&
0.924&0.891 \\\hline 30 &21894 &8666 &0.776 &1 &0.930 & 0.900
\\\hline 31 & 7627& 3841& 0.685& 1&0.933 &0.930 \\\hline 32 & 4185&
2242& 0.675& 1&0.921&0.929 \\\hline 33 & 23822& 10753& 0.648& 1&
0.959&0.917 \\\hline 34 & 8829 & 40 & 3.0 &0.33 &0.34 & 0.35
\\\hline
35 & 237982 &56961 & 0.462& 1& 0.993& 0.929 \\
\hline \hline
\end{tabular}
\end{table}

\end{document}